\documentclass[epj]{svjour}
\usepackage{color,epsfig,rotating,graphicx,subfigure}
\usepackage{amsmath,amssymb,amscd}
\usepackage[latin1]{inputenc}
\usepackage[english]{babel}
\usepackage{dsfont}
\usepackage{textcomp}

\renewcommand{\Re}{{\rm Re}}
\renewcommand{\Im}{{\rm Im}}

\newcommand{\ri}{{\rm i}}
\newcommand{\re}{{\rm e}}
\newcommand{\rd}{{\rm d}}

\newcommand{\kb}{k_{\rm B}}

\newcommand{\rpr}{{\rm pr}}
\newcommand{\rev}{{\rm ev}}

\newcommand{\rE}{{\rm E}}
\newcommand{\rH}{{\rm H}}  

\newcommand{\rS}{{\rm S}}
\newcommand{\rP}{{\rm P}}  

\newcommand{\rs}{{\rm s}}
\newcommand{\rp}{{\rm p}}
\newcommand{\rt}{{\rm t}}  

\newcommand{\red}{{\rm ed}}  
\newcommand{\rf}{{\rm f}}  

\begin{document}

\title{Shape-dependence of near-field heat transfer between a spheroidal 
        nanoparticle and a flat surface}

\author{Oliver Huth 
	\and Felix R\"uting 
	\and Svend-Age Biehs\thanks{Present address: 
		Laboratoire Charles Fabry,  
		Institut d'Optique, CNRS, Universit\'{e} Paris-Sud, 
		Campus Polytechnique, RD128, 91127 Palaiseau cedex, France} 
	\and Martin Holthaus}

\institute{Institut f\"ur Physik, Carl von Ossietzky Universit\"at,
   	D-26111 Oldenburg, Germany}

\date{January 12, 2010}

\abstract{
We study the radiative heat transfer between a spheroidal metallic nanoparticle
and a planar metallic sample for near- and far-field distances. In particular,
we investigate the shape dependence of the heat transfer in the near-field 
regime. In comparison with spherical particles, the heat transfer typically 
varies by factors between $1/2$ and $2$ when the particle is deformed such that 
its volume is kept constant. These estimates help to quantify the deviation of 
the actual heat transfer recorded by a near-field scanning thermal microscope
from the value provided by a dipole model which assumes a perfectly spherical 
sensor.
}

\titlerunning{Shape-dependence of near-field heat transfer}

\maketitle


\section{Introduction}

The optical properties of metallic nanoparticles depend significantly on 
their shapes, as has been demonstrated, e.g., for elliptical gold particles
attached to the apex of fiber-based probes for near-field optical 
microscopy~\cite{SqalliEtAl02}. Similarly, the thermal near-field radiation 
exchanged between a nanometer-sized particle at temperature $T_{\rP}$ and a 
closely spaced surface at different temperature $T_{\rS}$ is influenced by 
the particle's shape, although its linear extension may be significantly 
smaller than the dominant thermal wavelength. This shape-dependence occurs 
even when the particle's volume, and thus the total amount of radiating matter,
is kept constant. In this paper we quantify this effect for the case of 
spheroidal metallic nanoparticles.

Thermally induced near fields have attracted much experimental and theoretical 
attention in last years~\cite{JoulainEtAl05,VolokitinPersson07}. Several 
experiments and models have been designed in order to measure and describe the 
radiative heat transfer in the near-field regime, i.e., for distances smaller 
than the thermal wavelength. In this regime tunneling of thermal photons leads 
to a magnitude of heat transfer which can exceed that achieved by black-body 
far-field radiation by several orders of magnitude. Possible applications 
of this phenomenon include, in particular, thermophotovoltaic 
devices~\cite{DiMatteoEtAl01,NarayanaswamyChen03,MLarocheEtAl06,FrancoeurEtAl08,BasuEtAl09}.  

The near-field radiative heat transfer between two dielectric bodies 
has been calculated within the framework of Rytov's fluctuational 
electrodynamics~\cite{RytovEtAl89} for various geometries, 
including two semi-infinite planar bodies and layered structures (see, 
e.g.,~\cite{PolderVanHove71,LevinEtAl80,LoomisMaris94,JPendry99,Pan00,VolokitinPersson04,Bimonte06,Biehs07}), 
two spheres (e.g.,~\cite{NaChen08,ChapuisEtAl08b,PerezMadrid08}), 
a sphere above a semi-infinite body
(e.g.~\cite{JPendry99,Dorofeyev98,MuletEtAl01,DedkovKyasov07,ChapuisEtAl08,VoPer01}), 
and certain other two-di\-mensional geometries~\cite{DorofeyevEtAl99}. 
Early experiments to detect the radiative heat transfer between two 
effectively semi-infinite bodies with flat surfaces have been performed 
by Hargreaves~\cite{Hargreaves69}, and by 
Domoto {\itshape et al.}~\cite{DomotoEtAl70} Also a pioneering, but 
unconclusive experiment~\cite{XuEtAl94} by Xu {\itshape et al.} needs 
to be mentioned. An accurate measurement using glass plates separated 
by a micron-sized gap has been reported only recently by 
Hu {\itshape et al.}~\cite{HuEtAl08} Moreover, there now exist 
experimental setups measuring the near-field radiative heat transfer between 
a sphere and a flat sample~\cite{NarayaEtAl08,ShenEtAl09,RousseauEtAl09},
involving spheres with radii of some 10~\textmu m.

Relying on the same basic principle, a near-field scanning thermal microscope 
(NSThM) has been developed by Kittel 
{\itshape et al.}~\cite{MuellerHirsch99,KittelEtAl05,KittelEtAl08} for 
recording the radiative heat transfer at probe-sample distances even down to 
some nanometers. The sensor of this device consists of the tip of a scanning 
tunneling microscope, functionalized to act as a thermocouple, so that here 
the sensor-sample geometry differs significiantly from those geometries for 
which the radiative heat transfer has been calculated exactly. The foremost 
part of such a tip typically has a radius of less than 50~nm.

If one regards this active part of the NSThM sensor as a sphere, one may 
employ the familiar dipole model (see, 
e.g.,~\cite{JPendry99,Dorofeyev98,MuletEtAl01,DedkovKyasov07,ChapuisEtAl08}) 
for describing the results obtained with such an instrument quantitatively. 
However, the actual shape of the sensor is somewhat prolonged, resembling 
more an ellipsoid than a sphere, and varies from specimen to specimen, due to 
the difficult production process. Such variations of the geometry will have 
consequences for the signal measured with the NSThM. This is what motivates 
our present investigation of the shape-dependence of the near-field radiative 
heat transfer: In this work, we study the change of the near-field radiative 
heat transfer in response to shape variations of ellipsoidal nanoparticles 
above a flat sample. Besides helping one to estimate errors implied by the 
use of the dipole model, there may also be other, more general 
nanotechnological applications.  

The paper is organized as follows: In Sec.~\ref{Sec_dipolemodel} we briefly 
review the dipole model for calculating the near-field heat transfer between 
a dielectric sphere and a planar sample. This step mainly serves to collect 
the required material in a form which can be easily generalized. In 
Sec.~\ref{Sec_Dipole_Elli} we extend this dipole model to the geometry of an 
ellipsoidal particle above a flat sample. The near-field heat transfer for 
this geometry is then calculated explicitly in Sec.~\ref{Sec_Heat_Transfer}. 
In Sec.~\ref{Sec_Discus} we discuss the shape dependence of the heat transfer 
as the axes of the ellipsoid are varied, and compare our results to the heat 
transfer between a flat sample and a sphere.

\section{Dipole model}
\label{Sec_dipolemodel}

The near-field heat transfer between a spherical particle with temperature 
$T_{\rm \rP}$ and a sample with a planar surface and temperature $T_{\rm \rS}$ 
can be estimated by means of a simplified dipole model developed previously 
by several 
authors (e.g.,~\cite{JoulainEtAl05,VolokitinPersson07,JPendry99,Dorofeyev98,MuletEtAl01,DedkovKyasov07,ChapuisEtAl08}). 
To this end, one first considers the thermally fluctuating 
electric and magnetic fields $\mathbf{E}_{\rm f}$ and $\mathbf{H}_{\rm f}$ 
outside the sample, being generated by thermally fluctuating charges in its 
interior. These fluctuating fields consist of a propagating and of an 
evanescent part. The heat flux radiated by the sample is given by the mean 
Poynting vector $\langle \mathbf{E}_{\rm f} \times \mathbf{H}_{\rm f} \rangle$ 
and yields the well known Kirchhoff-Planck radiation law, to which only the 
propagating modes contribute, whereas the evanescent modes do not figure here, 
since they are bound to the surface of the sample. 

\begin{figure}[Hhbt]
\epsfig{file=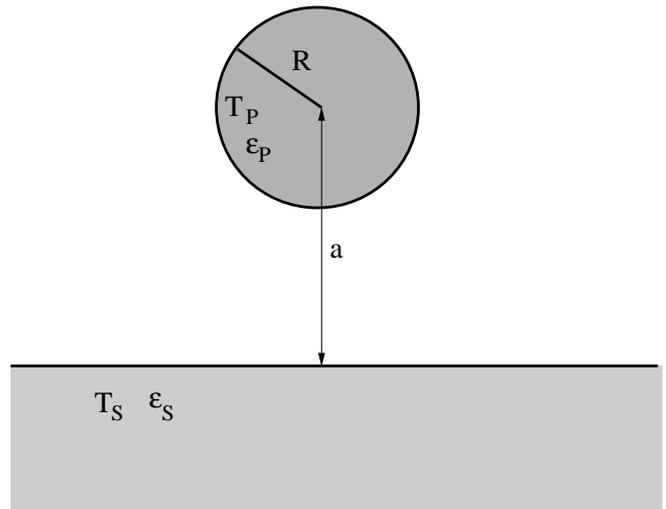, width=0.48\textwidth}
\caption{Sketch of a spherical nanoparticle with radius~$R$, 
	temperature~$T_{\rP}$, and permittivity~$\varepsilon_{\rP}$ 
	placed at a distance~$a$ above a planar sample with 
	temperature~$T_{\rS}$ and permittivity~$\varepsilon_{\rS}$.}
\label{Fig:Dipolmodell}
\end{figure}

Now, if an additional spherical particle with radius~$R$ is placed at a 
distance~$a$ above the sample, as depicted in Fig.~\ref{Fig:Dipolmodell}, 
the fluctuating fields $\mathbf{E}_{\rm f}$ and $\mathbf{H}_{\rm f}$ induce 
an electric dipole moment $\mathbf{p}$ and a Foucault current 
$\mathbf{j}_{\rm ed}$ within the particle. As emphasized by Chapuis
{\itshape et al.}~\cite{ChapuisEtAl08}, this eddy current 
$\mathbf{j}_{\rm ed}$ is particularly important for metallic particles. 
It causes losses inside the particle which can be described by means of an 
effective magnetic dipole moment $\mathbf{m}_{\rm eff}$, in analogy to the 
electric dipole moment $\mathbf{p}$. We assume $a \gg R$, so that higher 
multipoles may be neglected; in principle, this restriction to sufficiently
large distances~$a$ could be removed by including higher multipoles. Within
the dipole approximation, the energy absorbed by the particle can then be 
written as
\begin{equation}
	\langle P_{\rS \rightarrow \rP}\rangle = 
	\langle\dot{\mathbf{p}} \cdot \mathbf{E}_{\rm f}\rangle
	+ \langle\dot{\mathbf{m}}_{\rm eff} \cdot \mathbf{H}_{\rm f}\rangle \; ;
\label{Eq:P_Sample_to_Probe_general}
\end{equation}
the angular brackets denote an ensemble average. For a homogenous and isotropic 
particle the relations between the induced dipole moments and the fields are 
given by
\begin{equation}
  	\mathbf{p} = \varepsilon_0 \, \mathbf{\alpha}_{\rm p}^{\rm E} 
	\mathbf{E}_{\rm f}
\label{Eq:Dipole_moment_el}
\end{equation}
and
\begin{equation}
	\mathbf{m}_{\rm eff} = \mu_0 \, \mathbf{\alpha}_{\rm p}^{\rm H} 
	\mathbf{H}_{\rm f} \; ,
\label{Eq:Dipole_moment_ma}
\end{equation}
where $\mathbf{\alpha}_{\rm p}^{\rm E}$ and $\mathbf{\alpha}_{\rm p}^{\rm H}$ 
symbolize the electric and magnetic polarizabilities of the particle; 
$\varepsilon_0$ and $\mu_0$ are the permittivity and the permeability of the 
vacuum, respectively. In general, the polarizabilities have a directional 
dependence and therefore are described by a tensor. For the highly symmetric 
case of a sphere, this tensor reduces to a scalar multiple of the unit tensor, 
so that the polarizabilities are represented by scalar values in this case. 
It is essential to keep in mind that only non-magnetic materials are taken 
into account here, so that the absorption 
$\langle\dot{\mathbf{m}}_{\rm eff} \cdot \mathbf{H}_{\rm f}\rangle$ 
is solely ascribed to the loss due to eddy currents. 

With the help of Eqs.~(\ref{Eq:Dipole_moment_el}) and 
(\ref{Eq:Dipole_moment_ma}) one obtains the expression
\begin{equation}
\begin{split}
 	\langle P_{\rS \rightarrow \rP}\rangle (T_\rS) 
	&= \int_0^\infty \! \rd \omega \, \omega \left[
	\Im(\alpha_{\rP}^{\rm E})\varepsilon_0
        \langle\left|\mathbf{E}_{\rm f}\right|^2\rangle_\omega \right.\\
	&\qquad\left.+\Im(\alpha_{\rP}^{\rm H}) \mu_0 
	\langle\left|\mathbf{H}_{\rm f} \right|^2\rangle_\omega\right]
\label{Eq:P_Sample_to_Probe_polarisablility}
\end{split}
\end{equation}
for the energy transferred from the sample to the 
particle~\cite{ChapuisEtAl08,Dorofeyev08}. 
Here $\langle\left|\mathbf{E}_{\rm f}\right|^2\rangle_\omega$ denotes the 
frequency-dependent correlation function of the electric field, i.e.,
the expectation value of the product of the Fourier transform 
$\mathbf{E}_{\rm f}(\omega)$ of the fluctuating electric field and 
its complex conjugate; the correlation function 
$\langle\left|\mathbf{H}_{\rm f} \right|^2\rangle_\omega$ of the magnetic 
field is defined analogously.  

Reversely, fluctuating charges inside the sphere of temperature $T_\rP$ 
give rise to energy transfer from the sphere to the flat surface. 
The power absorbed within the sample, denoted 
$\langle P_{\rP \rightarrow \rS}\rangle(T_\rp)$, 
can be calculated in the same manner as 
$\langle P_{\rS\rightarrow \rP}\rangle(T_\rS)$ 
above. The  resulting expression differs from the 
expression~(\ref{Eq:P_Sample_to_Probe_polarisablility}) only through the 
temperature, so that $\langle P_{\rP \rightarrow \rS}\rangle(T_\rp) 
= \langle P_{\rS \rightarrow \rP}\rangle(T_\rP)$. As the energy flux 
$\langle P_{\rP\rightarrow \rS}\rangle$ is directed oppositely to 
$\langle P_{\rS\rightarrow \rP}\rangle$, the resulting overall heat 
transfer between the particle and the sample has the 
form~\cite{ChapuisEtAl08,Dorofeyev08}
\begin{equation}
	\langle P_{\rS \leftrightarrow \rP}\rangle =
	\langle P_{\rS \rightarrow \rP}\rangle(T_\rS) - 
     	\langle P_{\rS \rightarrow \rP}\rangle(T_\rP) \; .
\label{Eq:P_Sample_Probe_polarisablility}
\end{equation}
The correlation functions 
$\langle\left|\mathbf{E}_{\rm f}\right|^2\rangle_\omega$ and
$\langle\left|\mathbf{H}_{\rm f}\right|^2\rangle_\omega$ in 
Eq.~(\ref{Eq:P_Sample_to_Probe_polarisablility}) are known from fluctuational 
electrodynamics~\cite{RytovEtAl89}. If one assumes that the sample occupies
the infinite half space $z \le 0$, as in Ref.~\cite{JoulainEtAl05}, 
one gets  
\begin{equation}
\begin{split}
    &	\langle E_{{\rm f},i}\,E_{{\rm f},i}^*\rangle_\omega 
      = \Theta(\omega, T_{\rm S})\frac{2\omega}{\pi\varepsilon_0 c^2}\biggl\{\\
    &   \qquad \int\limits_{\kappa<k_0} \!\!\!\frac{\rd^2 \kappa}{(2 \pi)^2}
    	\frac{1}{4 k_{z0}}
	\biggl[\bigl(\mathbf{e}_\perp \otimes \mathbf{e}_\perp\bigr)_{ii}
	\bigl(1 - |r_\perp|^2\bigr)\\ 
    &	\qquad\qquad\qquad\quad\,\,\,\,+ \bigl(
	\mathbf{e}_\parallel^{\rm pr} \otimes \mathbf{e}_\parallel^{\rm pr}
	\bigr)_{ii} \bigl(1 -|r_\parallel|^2\bigr) \biggr] \\
    &   \quad +\int\limits_{\kappa>k_0} \!\!\! \frac{\rd^2 \kappa}{(2\pi)^2}
        \frac{\re^{-2\gamma\,a}}{2\gamma}
	\biggl[\Im(r_\perp) \,
	\bigl(\mathbf{e}_\perp \otimes \mathbf{e}_\perp\bigr)_{ii}\\
    &  \qquad\qquad\qquad\qquad  + \Im(r_\parallel) \bigl(
        \mathbf{e}_\parallel^{\rm ev} \otimes \mathbf{e}_\parallel^{\rm ev}
	\bigr)_{ii}\biggr]\biggr\}
\end{split}
\label{Eq:Korrelation_EE_halfspace}
\end{equation}
for the electric correlation function, where
\begin{equation}
  	\Theta(\omega, T) = \frac{\hbar\omega}{\re^{\hbar\omega\beta} -1} 
\label{Eq:Bose_Einstein}
\end{equation}
is the Bose function with the inverse temperature $\beta = 1 /(\kb T)$. 
Moreover, the unit vectors
\begin{equation}
\begin{split}
  	\mathbf{e}_\perp & := \frac{1}{\kappa}\left(-k_y, k_x, 0\right)^\rt \\
  	\mathbf{e}_{\parallel}^{\rm pr} & := \frac{1}{k_0\kappa}
	\left(k_x k_{z0}, k_y k_{z0}, \kappa^2\right)^\rt \\
  	\mathbf{e}_{\parallel}^{\rm ev} & := \frac{1}{k_0\kappa}
	\left(k_x \gamma, k_y \gamma, \kappa^2\right)^\rt 
\end{split}
\label{Eq:Def_TE_TM}
\end{equation}
have been introduced, writing $\kappa^2 =k_x^2 + k_y^2$ together with 
$k_{z0} = \sqrt{k_0^2 - \kappa^2}$ and $\gamma = \sqrt{\kappa^2 - k_0^2}$.
As usual, $k_0 = \omega/c$, where $c$ is the velocity of light in vacuum. 
In addition, we write 
\begin{equation}
\begin{split}
  	r_\perp &= \frac{k_{z0}-k_z}{k_{z0}+k_z} \\ 
  	r_\parallel &= \frac{\varepsilon_{\rm S} k_{z0}-k_z}
	{\varepsilon_{\rm S} k_{z0}+k_z}
\end{split}
\end{equation}
for the Fresnel reflection coefficients, with $k_z=\sqrt{\varepsilon k_0^2 - \kappa^2}$. 
The first integral in 
Eq.~(\ref{Eq:Korrelation_EE_halfspace}), with $\kappa < k_0$, describes the 
propagating part of the fluctuating field, whereas the second integral with 
$\kappa > k_0$ describes the evanescent part. The expression for 
$\langle H_{{\rm f},i}H_{{\rm f},i}^*\rangle_\omega$ is obtained from 
Eq.~(\ref{Eq:Korrelation_EE_halfspace}) by interchanging 
$r_\perp \leftrightarrow r_\parallel$ and 
$\frac{1}{\varepsilon_0} \leftrightarrow \frac{1}{\mu_0}$, due to a 
corresponding symmetry of the electric and magnetic Green's functions.

Knowing the correlation functions above a half space, the mean power 
absorbed by the particle above a planar surface can be calculated from 
Eq.~(\ref{Eq:P_Sample_to_Probe_polarisablility}), provided the polarizabilities 
$\alpha_{\rm P}^{\rm E}$ and $\alpha_{\rm P}^{\rm H}$ are given. In the case 
of a spherical particle of radius~$R$ these quantities can be derived from 
Mie scattering theory~\cite{BohrHuff}. Denoting the particle's relative
permittivity as $\varepsilon_{\rm P}$, and introducing the dimensionless
variables $x = k_0 R$ and $y = \sqrt{\varepsilon_\rP} k_0 R$, 
one finds~\cite{ChapuisEtAl08b,BohrHuff} 
\begin{equation}
\begin{split}
	\alpha_{\rm P}^{\rm E} & = 
	2\pi R^3 \frac{\left(2 \varepsilon_{\rm P} + 1\right)\left(\sin(y) - y \cos(y)\right)
	                -y^2\sin(y)}
		       {\left(\varepsilon_{\rm P} - 1\right)\left(\sin(y) - y \cos(y)\right)
	                +y^2\sin(y)} \\
    	\alpha_{\rm P}^{\rm H} & = 
	\frac{\pi R^3}{3}\left[\frac{\left(x^2-6\right)}{y^2}\!\!\left(y^2 + 3 y \cot(y) - 3\right) 
	   - \frac{2 x^2}{5}\right] 
\end{split}
\end{equation}
for $x \ll 1$, implying that the particle's radius should be small compared
to the dominant thermal wavelength, $R \ll \lambda_{\rm th}$. If we demand 
even $|y| \ll 1$, meaning that the radius be smaller than the skin depth
at thermal frequencies, the above expressions reduce to
\begin{align}
  	\alpha_{\rm P}^{\rm E} & = 
	4\pi\,R^3 \frac{\varepsilon_{\rm P}-1}{\varepsilon_{\rm P}+2} \; , 
\label{Eq:Clausius-Mossotti}\\
  	\alpha_{\rm P}^{\rm H} & = 
	\frac{2\pi}{15} R^3\left(k_0 R\right)^2
	\left(\varepsilon_{\rm P}-1\right) \; .
  \label{Eq:Eddy_currrents}
\end{align}
Evidently, the expression for $\alpha_{\rm P}^{\rm E}$ equals the well known 
Clausius-Mossotti formula. 

Before proceeding, we specify the relevant orders of magnitude. For bulk
metals the relative permittivity is well described by the Drude 
model~\cite{AshcroftMermin76}
\begin{equation}
  	\varepsilon = 1 - \frac{\omega_p^2}
	{\omega\left(\omega+\ri\omega_{\tau}\right)} \; .
\label{Eq:Drude-Modell}
\end{equation}
For gold at room temperature the plasma frequency is
$\omega_p = 1.4 \times 10^{16} \; \text{s}^{-1}$, 
while the relaxation rate figures as 
$\omega_\tau = 3.3 \times 10^{13} \; \text{s}^{-1}$. Taking the thermal 
frequency $\omega_{\rm th} \approx 10^{14} \; \text{s}^{-1}$ one then finds
$(\sqrt{|\epsilon|} k_0)^{-1} \approx 21$~nm, setting a rough upper limit to the 
radius of a gold sphere for which the expressions~(\ref{Eq:Clausius-Mossotti}) 
and (\ref{Eq:Eddy_currrents}) may still hold with reasonable accuracy. 
When treating such minuscule particles we have to modify the Drude 
permittivity~(\ref{Eq:Drude-Modell}) in order to account for surface 
scattering, since the bulk mean free path of the electrons, which is about 
$42$~nm for gold, reaches the same order of magnitude as the spheres' 
diameters. For spherical particles of radius~$R$ the required correction is 
achieved by the replacement~\cite{TomGri} 
\begin{equation}
	\omega_\tau \to \widetilde{\omega}_\tau 
	= \frac{3}{4}\frac{v_F}{R} \; ,
\label{Eq:SurfScat}
\end{equation}
where $v_F$ is the Fermi velocity. Taking $v_F = 1.4 \times 10^6$~m/s
for gold, and setting $R = 20$~nm, we obtain 
$\widetilde{\omega}_\tau = 5.3 \times 10^{13} \; \text{s}^{-1}$, less than
two times the bulk value. On the other hand, size quantization effects become 
essential only for radii below 2~nm, and may therefore be neglected here.

\section{Dipole model for ellipsoidally shaped particles}
\label{Sec_Dipole_Elli}

\begin{figure}[Hhbt]
\epsfig{file=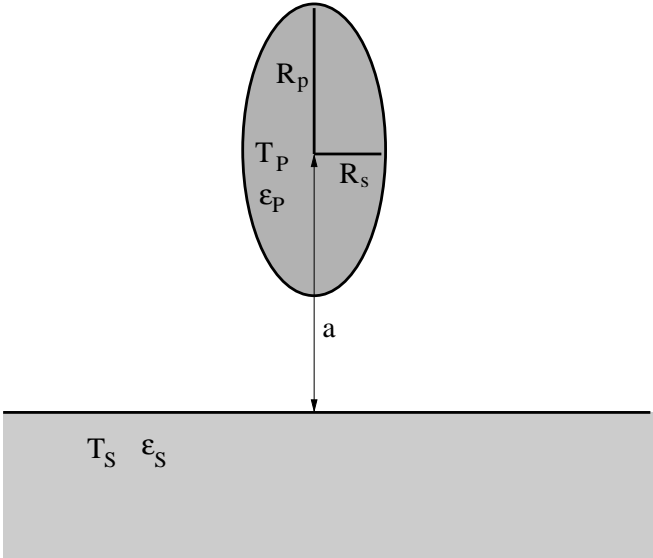, width=0.48\textwidth}
\caption{Sketch of a spheroidally shaped particle with semi-axes  $R_{\rs}$ 
	and $R_{\rp}$. The semi-axis with length $R_{\rp}$ lies along the
	spheroid's axis of rotation.} 
\label{Fig:Ellipsoid}
\end{figure}

The near-field radiative heat transfer between a flat sample with temperature $T_\rS$ 
and an ellipsoidally shaped particle with temperature $T_\rP$, as sketched in 
Fig.~\ref{Fig:Ellipsoid}, can be expressed through equations similar to 
Eqs.~(\ref{Eq:P_Sample_to_Probe_polarisablility}) 
and (\ref{Eq:P_Sample_Probe_polarisablility}), but instead of employing the 
expressions $\mathbf{\alpha}_{\rP}^{\rE/\rH}$ for the polarizabilities of a 
sphere, the proper polarizability tensors for an ellipsoidal particle have 
to be taken into account.

Denoting the three semi-axes as $R_x$, $R_y$, $R_z$, we require 
$a \gg \max\{R_x, R_y, R_z\} \equiv R_{\max}$ to substantiate the dipole 
approximation, while keeping $R_{\max}$ smaller than the skin depth, thus 
allowing for approximately homogeneous fields inside the ellipsoids. It is 
textbook knowledge that the induced dipole moment $\mathbf{p}$ of an 
ellipsoidal particle is connected with the incident field $\mathbf{E}_\rf$ 
through the relation~\cite{LandLif}
\begin{equation}
  	p_i = \varepsilon_0\, V_\rE 
	\frac{\varepsilon_\rP - 1}{1 + (\varepsilon_\rP - 1) n_i} E_{\rf,i}
\label{Eq:elec_dipole_E}
\end{equation}
with $V_\rE= \frac{4 \pi}{3} R_x R_y R_z$ specifying the volume of the 
particle, and $\varepsilon_\rP$ its permittivity. It is assumed here that the 
components of $\mathbf{p}$ and $\mathbf{E}_\rf$ are given in the principal-axis
system of the ellipsoid. The polarizability tensor then is diagonal, with
diagonal elements given by
\begin{equation}
  	\alpha_{\rE,i}^\rE = V_\rE 
	\frac{\varepsilon_\rP - 1}{1 + (\varepsilon_\rP - 1) n_i} \; .
\label{Eq:Polarisability_E}
\end{equation}
The quantities $n_i$ are the so-called depolarization coefficients. They 
are written as~\cite{LandLif}
\begin{equation}
  	n_i = \frac{R_x R_y R_z}{2} 
	\int_0^\infty\!\! \rd s \, \frac{1}{(s + R_i^2) R (s)}
\label{Eq:depolarization_factors}
\end{equation}
with $R (s) = \sqrt{(s + R_x^2)(s + R_y^2)(s + R_z^2)}$. These coefficients 
depend only on the shape of the ellipsoid, not on its  volume. They obey 
the relations
\begin{equation}
  	\sum_{i = 1}^{3} n_i = 1 \quad \text{and} \quad n_i \geq 0 \; .
\end{equation}
For the special case of a sphere with $R_x = R_y = R_z \equiv R$, 
Eq.~(\ref{Eq:depolarization_factors}) yields $n_1 = n_2 = n_3 = 1/3$, 
so that Eq.~(\ref{Eq:Polarisability_E}) correctly reduces to the 
Clausius-Mossotti formula~(\ref{Eq:Clausius-Mossotti}). In the case of 
spheroid, i.e., of a rotational ellipsoid with two equal semi-axes, 
$R_x = R_y \equiv R_\rs$ and $R_z \equiv R_\rp$, the coefficients $n_i$ 
take the form~\cite{LandLif}
\begin{equation}
\begin{split}
  	n_1(e) &= n_2(e) = \frac{1}{2} \bigl( 1 - n_3(e) \bigr),\\
  	n_3(e) & = 
	\begin{cases} 
              \frac{1 - e^2}{2 e^3} \biggl(\ln\biggl(\frac{1+e}{1-e}\biggr) 
	      - 2 e\biggr) \; , & R_\rs < R_\rp  \\ 
              \frac{1 + e^2}{e^3} \bigl( e - \arctan(e) \bigr) \; , & 
	      R_\rs > R_\rp \; ,
	\end{cases} 
\end{split}
\label{Eq:depolarization_factors_rotE}
\end{equation}
with
\begin{equation}
  	e^2 = \begin{cases}
		1 - \frac{R_\rs^2}{R_\rp^2} \; , & R_\rs < R_\rp \\
           	\frac{R_\rs^2}{R_\rp^2} - 1 \; , & R_\rs > R_\rp \; . 
        \end{cases} 
\label{Eq:Hilfsgroesse}
\end{equation}

For calculating the effective magnetic polarizability of an ellipsoid,
we follow a strategy outlined by Tomchuk and Grigorchuk~\cite{TomGri},
and use the identity 
\begin{equation}
  	\omega \, \Im(\alpha_{\rE,i}^{\rm H}) \mu_0 
	\langle\left|H_{\rm f,i}\right|^2\rangle_\omega  = 
        \Re\!\! \int_{V_\rE}\!\!\! \! \rd^3 r \, 
	\langle \mathbf{j}_\red \cdot \mathbf{E_{\rf,\red}}^* \rangle_\omega
	\; ,
\label{Eq:absorption_E}
\end{equation}
thus shifting the emphasis from the effective magnetic moment 
$\mathbf{m}_{\rm eff}$ to a fluctuating eddy field $\mathbf{E_{\rf,\red}}$. 
This requires that the material is non-magnetic, so that $\mathbf{m}_{\rm eff}$
is entirely due to eddy currents $\mathbf{j}_\red$. The eddy field obeys the 
equations 
\begin{align}
  	\nabla \times \mathbf{E}_{\rf,\red}(\omega) & = 
	\ri \omega \mu_0 \mathbf{H}_\rf(\omega) \; , \\
  	\nabla \cdot \mathbf{E}_{\rf,\red}(\omega)  & = 0 \; ;
\end{align}
in principle, the second equation constitutes a boundary condition for 
$\mathbf{E}_{\rf,\red}$. Because we assume $R_{\max}$ to be smaller than 
the skin depth, $\mathbf{H}_\rf$ can be considered as constant within the 
volume of the particle, so that the eddy field $\mathbf{E}_{\rf,\red}$ 
depends linearly on the spatial coordinates. The resulting relation between 
$\mathbf{E}_{\rf,\red}$ and $\mathbf{H}_\rf$ reads~\cite{TomGri}
\begin{equation}
  	E_{{\rm f},\red,x}(\omega) = 
	\ri \omega \mu_0 R_x^2\biggl( 
	\frac{z H_{\rf,y}(\omega)}{R_z^2 + R_x^2} - 
        \frac{y H_{\rf,z}(\omega)}{R_x^2 + R_y^2} \biggr) \; . 
\label{Eq:E_H_Ellipsoid}
\end{equation}
The other components of the electric eddy field are obtained by cyclic 
permutation of the indices. We emphasize that the underlying assumption of 
constant fields $\mathbf{H}_\rf$ within the particle is not well fulfilled 
in cases where the particle's characteristic linear dimensions are on the 
order of the skin depth, which is about 21~nm for gold at room temperature. 
In such cases, the formalism outlined here may still yield the correct orders 
of magnitude, but not exact numbers.   

Next, one has the identity~\cite{TomGri}
\begin{equation}
  	\mathbf{j}_\red(\omega) = 
	\omega \varepsilon_0 \varepsilon_\rP'' \mathbf{E}_{\rf,\red}(\omega)
	\; ,
\end{equation}
which relates the induced eddy currents to $\mathbf{E}_{\rf,\red}$, with 
$\varepsilon_\rP''$ denoting the imaginary part of the particle's 
permittivity. Thus, for a spheroid the absorbed energy~(\ref{Eq:absorption_E}) 
is found to be
\begin{equation}
\begin{split}
      &	\Re\int_{V_\rE} \! \rd^3 r \, 
      	\langle \mathbf{j}_\red \cdot \mathbf{E_{\rf,\red}}^* \rangle_\omega
      = \frac{\omega^3}{c^2} \mu_0 \varepsilon_{\rP}'' \frac{V_\rE}{10} \times\\
      & \biggl[ \langle |H_{\rf,z}|^2 \rangle_\omega R_\rs^2 
        + \langle |H_{\rf,x}|^2 + |H_{\rf,y}|^2 \rangle_\omega 
	\frac{2 R_\rs^2 R_\rp^2}{R_\rs^2 + R_\rp^2}\biggr] \; .     
\end{split}
\end{equation}
The effective polarizability is finally read off by comparing this expression 
with the second term in Eq.~(\ref{Eq:P_Sample_to_Probe_polarisablility}). 
Again the polarizability tensor has non-zero entries only on its diagonal, 
with imaginary parts given by
\begin{equation}
\begin{split}
  	\Im(\alpha_{\rE,x}^\rH) & = 
	\Im(\alpha_{\rE,y}^\rH) = 
	\frac{(k_0 R_\rs)^2 R_\rp^2}{R_\rs^2 + R_\rp^2} 
	\frac{V_\rE \varepsilon_\rP''}{5} \; ,\\
  	\Im(\alpha_{\rE,z}^\rH) & = 
	\frac{ (k_0 R_\rs)^2}{10} V_\rE \varepsilon_\rP'' \; . 
\end{split}
\label{Eq:magn_polarisability_E}
\end{equation}
For a sphere with $R_\rs = R_\rp = R$, both expressions give the correct 
imaginary part of the Mie formula~(\ref{Eq:Eddy_currrents}).

\begin{figure}[Hhbt]
\epsfig{file=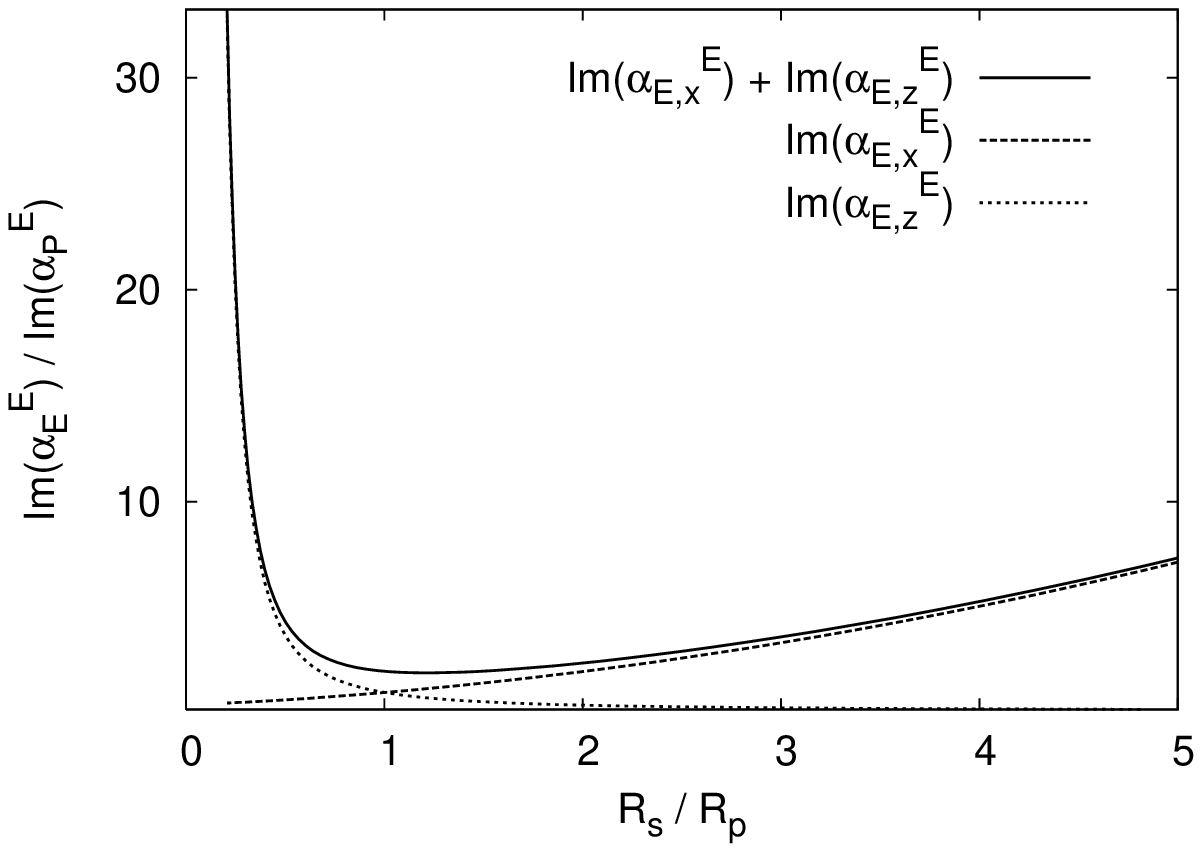, width=0.48\textwidth}
\caption{Imaginary parts $\Im(\alpha_{\rE,x}^\rE)$ and $\Im(\alpha_{\rE,z}^\rE)$
	of the electric polarizabilities~(\ref{Eq:Polarisability_E}) at 
	$\omega_{\rm th} = 10^{14} \; \text{s}^{-1}$ for a spheroidal gold 
	particle, and their sum, vs.\ the ratio $R_\rs/R_\rp$. The spheroid's 
	volume $V_{\rm E}$ is kept constant at that of a sphere with radius 
	$R = 20$~nm; surface scattering is taken into account. Data are 
	normalized by the imaginary part of the electric polarizability 
	$\Im(\alpha_\rP^\rE)$ of the sphere.}
\label{Fig:Alpha_el}
\end{figure}

\begin{figure}[Hhbt]
\epsfig{file=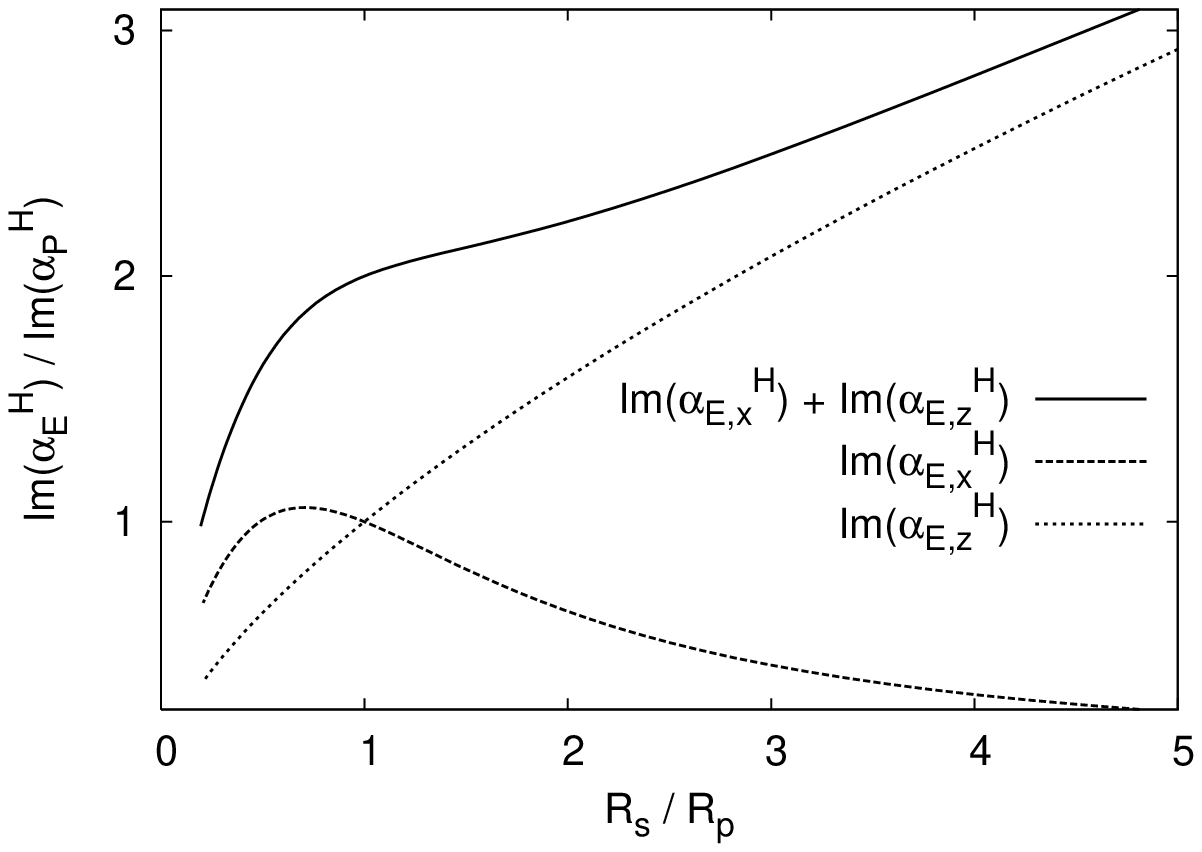, width=0.48\textwidth}
\caption{Imaginary parts $\Im(\alpha_{\rE,x}^\rH)$ and $\Im(\alpha_{\rE,z}^\rH)$
	of the magnetic polarizabilities~(\ref{Eq:magn_polarisability_E}) at 
	$\omega_{\rm th} = 10^{14} \; \text{s}^{-1}$ for a spheroidal gold 
	particle, and their sum, vs.\ the ratio $R_\rs/R_\rp$. The spheroid's 
	volume $V_{\rm E}$ is kept constant at that of a sphere with radius 
	$R = 20$~nm; surface scattering is taken into account. Data are 
	normalized by the imaginary part of the magnetic polarizability 
	$\Im(\alpha_\rP^\rH)$ of the sphere.}
\label{Fig:Alpha_mg}
\end{figure}

\begin{figure}[Hhbt]
\epsfig{file=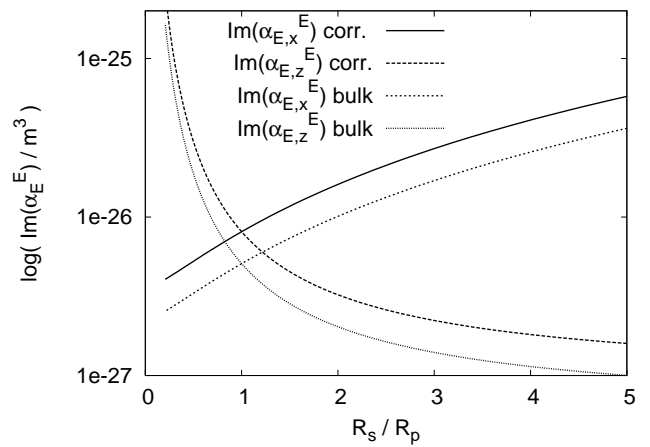, width=0.48\textwidth}
\caption{Electric absorptivities for spheroidal gold particles with
	a fixed volume corresponding to that of a sphere with radius
	$R = 20$~nm, both with (corrected) and without (bulk data) surface 
	scattering taken into account.} 
\label{Fig:Comp}
\end{figure}

In Figs.~\ref{Fig:Alpha_el} and \ref{Fig:Alpha_mg} we plot the imaginary parts 
of the electric and of the magnetic polarizability~(\ref{Eq:Polarisability_E})
and (\ref{Eq:magn_polarisability_E}) at the thermal frequency
$\omega_{\rm th} = 10^{14} \; \text{s}^{-1}$ against the ratio $R_\rs/R_\rp$, 
keeping the volume $V_{\rm E}$ constant at that of a sphere with radius 
$R = 20$~nm. The Drude expression~(\ref{Eq:Drude-Modell}) has been taken for 
the permittivity, with the plasma frequency 
$\omega_p = 1.4 \times 10^{16} \; \text{s}^{-1}$ for gold at room temperature, 
while surface scattering has been taken into account through the 
replacement~(\ref{Eq:SurfScat}), inserting the geometric mean
$(R_\rs^2 R_\rp)^{1/3}$ for $R$. This replacement is not fully correct for 
nonspherical particles~\cite{TomGri}: In principle, one then has different
scattering times for different directions, leading to anisotropy of the
permittivity. In order to estimate the resulting error, we compare in 
Fig.~\ref{Fig:Comp} the electric absorptivities provided by the Drude 
permittivity with the simple surface correction~(\ref{Eq:SurfScat}),
and without any correction at all. While the effect of the correction is 
clearly visible, the general trends are not altered. Hence, a more refined 
correction taking anisotropy into account should not give drastically
different results, at least not in the interval $1/5 \le R_\rs/R_\rp \le 5$
considered.

The curve progressions observed here can be understood intuitively. For the 
electric polarizability depicted in Fig.~\ref{Fig:Alpha_el} the absorptivity 
$\Im(\alpha_{\rE,z}^{\rE})$ decreases monotonically with increasing ratio 
$R_\rs/R_\rp$, whereas the absorptivity $\Im(\alpha_{\rE,x}^{\rE})$ increases. 
For $R_\rs/R_\rp \ll 1$ the spheroidal particle becomes needle-like along the 
$z$-direction; hence, the polarizability in $z$-direction is much greater 
than that in $x$- or $y$-direction. On the other hand, when $R_\rs/R_\rp \gg 1$
the spheroid is pancake-like and lies parallel to the $x$-$y$-plane; therefore 
the polarizability in the directions perpendicular to the $z$-direction 
dominates.

In the case of the magnetic polarizability shown in Fig.~\ref{Fig:Alpha_mg} 
one observes that both $\Im(\alpha_{\rE,x}^\rH)$ and $\Im(\alpha_{\rE,z}^\rH)$ 
go to zero in the formal limit $R_\rs/R_\rp \rightarrow 0$. This is clear 
because these two quantities represent absorptivities caused by eddy currents 
perpendicular to the $x$- and to the $z$-direction, respectively. Accordingly, 
for thin needle-like spheroids with $R_\rs/R_\rp \ll 1$ both absorptivities 
disappear, because no eddy current can arise then. In the opposite case, for 
$R_\rs/R_\rp \gg 1$, eddy currents can be easily induced perpendicular to the 
$z$-direction in the pancake-like particle, whereas perpendicular to the 
$x$-direction no eddy currents occur, as before. Hence, for 
$R_\rs/R_\rp \rightarrow \infty$ one expects 
$\Im(\alpha_{\rE,z}^\rH) \rightarrow \infty$ and 
$\Im(\alpha_{\rE,x}^\rH) \rightarrow 0$. Since $\Im(\alpha_{\rE,x}^\rH)$ 
is positive for all values of $R_\rs/R_\rp$ and goes to zero in both limiting 
cases there has to be a maximum, which occurs near $R_\rs/R_\rp = 1$, as 
witnessed by Fig.~\ref{Fig:Alpha_mg}.

\section{Radiative heat transfer between an ellipsoidal nanoparticle and a 
planar surface}
\label{Sec_Heat_Transfer}

In accordance with Sec.~\ref{Sec_dipolemodel}, the radiative heat transfer 
from a planar sample to an ellipsoidal nanoparticle is expressed, in 
analogy to Eq.\ (\ref{Eq:P_Sample_to_Probe_polarisablility}), as
\begin{equation}
\begin{split}
 	\langle P_{\rS\rightarrow \rE}\rangle(T_\rS) = 
	\int_0^\infty \!\!\! \rd \omega \,
	&\omega\left[
	\Im(\alpha_{\rE,i}^\rE)\varepsilon_0
        \langle \left|E_{{\rm f},i} \right|^2\rangle_\omega \right.\\
	&\left.+\Im(\alpha_{\rE,i}^\rH) \mu_0 
	\langle \left|H_{{\rm f},i} \right|^2\rangle_\omega
	\right] \; ,
\label{Eq:P_Sample_to_Probe_polarisablility_E}
\end{split}
\end{equation}
where summation over repeated indices is implied. It is assumed here that the 
particle is oriented as in Fig.~\ref{Fig:Ellipsoid}, so that the surface normal
is parallel to a principal axis of the ellipsoid. As discussed before, this 
form~(\ref{Eq:P_Sample_to_Probe_polarisablility_E}) is quite general, 
requiring only the specification of the polarizabilities of the nanoparticle, 
and of the correlation functions of the fluctuating fields above the 
sample's surface. The expression $\langle P_{\rE \rightarrow \rS}\rangle$ 
for the energy flux from the nanoparticle back to the sample can again be 
obtained from Eq.~(\ref{Eq:P_Sample_to_Probe_polarisablility_E}) by 
substituting the temperature $T_\rP$ for $T_\rS$.

When the specific expressions for a spheroid above a planar sample are inserted 
into Eq.~(\ref{Eq:P_Sample_to_Probe_polarisablility_E}), the propagating modes 
with $\kappa<k_0$ yield the contribution 
\begin{equation}
\begin{split}
      \langle P_{\rpr, \rS \rightarrow \rE}& \rangle  = 
	\frac{2}{\pi}\int_0^\infty \!\!\! \rd\omega\, \Theta(\omega,T_\rS) 
        k_0^2 \!\!\! 
	\int \limits_{\kappa < k_0} \!\!\!\frac{\rd^2 \kappa}{(2 \pi)^2} 
	\frac{1}{4 k_{z0}}\biggl\{ \\
      &  \Im(\alpha_{\rE,x}^\rE) \bigl[ (1 - |r_\perp|^2) 
	+ \frac{k_{z0}^2}{k_0^2}(1 - |r_\parallel|^2)\bigr] \\
      +&  \Im(\alpha_{\rE,z}^\rE) \frac{\kappa^2}{k_0^2} (1 - |r_\parallel|^2) 
	\\
      +&  \Im(\alpha_{\rE,z}^\rH)  \bigl[ (1 - |r_\parallel|^2) 
        + \frac{k_{z0}^2}{k_0^2}(1 - |r_\perp|^2)\bigr]\\
      +&  \Im(\alpha_{\rE,x}^\rH) \frac{\kappa^2}{k_0^2} (1 - |r_\perp|^2) 
	\biggr\} \; . 
\end{split}
\label{Eq:Sample_to_Ellipsoid_pr}
\end{equation} 
This reduces correctly to the known result for a sphere by setting 
$R_\rs = R_\rp = R$: 
\begin{equation}
\begin{split}
      &	\langle P_{\rpr, \rS \rightarrow \rP} \rangle  = 
	\frac{2}{\pi}\int_0^\infty \!\!\! \rd\omega\, \Theta(\omega,T_\rS) 
        k_0^2 
	\int\limits_{\kappa < k_0}\!\!\! \frac{\rd^2 \kappa}{(2 \pi)^2} 
	\frac{1}{4 k_{z0}} \\
      & \times\biggl\{ 
        \Im(\alpha_{\rP}^{\rE} + \alpha_{\rP}^{\rH}) 
	\bigl[ (1 - |r_\parallel|^2) + (1 - |r_\perp|^2)\bigr]\biggr\} \; . 
\end{split}
\label{Eq:Sample_to_Sphere_pr}
\end{equation}
Observe that sum of the polarizabilities figures here, because 
$\varepsilon_0\langle |\mathbf{E}_\rf|^2 \rangle_\omega = 
\mu_0\langle |\mathbf{H}_\rf|^2 \rangle_\omega$
for the propagating modes. Accordingly, the relative size of 
$\Im(\alpha_{\rP}^{\rE})$ and $\Im(\alpha_{\rP}^{\rH})$ determines whether 
the radiative heat transfer is dominated by the electric or by the magnetic 
part. From Eqs.~(\ref{Eq:Sample_to_Ellipsoid_pr}) 
and (\ref{Eq:Sample_to_Sphere_pr}) one can calculate the heat radiated by an 
ellipsoidal or a spherical nanoparticle in the absence of the 
sample~\cite{Martynenko2005} by simply setting $|r_\perp| = |r_\parallel| = 0$ 
and inserting the temperature $T_\rP$ instead of $T_\rS$. 

The heat transfer mediated by the evanescent modes with $\kappa > k_0$ from 
the sample to the spheroidal particle takes the form
\begin{equation}
\begin{split}
      	\langle P_{\rev, \rS \rightarrow \rE} \rangle  &= 
	\frac{2}{\pi}\int_0^\infty \!\!\!\rd\omega\,\Theta(\omega, T_\rS) 
         \,k_0^2\!\!\int\limits_{\kappa > k_0} \!\!\!\frac{\rd^2 \kappa}{(2 \pi)^2} 
	\frac{\re^{-2 \gamma a}}{2 \gamma} \\
      & \biggl\{ \Im(\alpha_{\rE,x}^\rE)
        \bigl[ \Im(r_\perp) + \frac{\gamma^2}{k_0^2} \Im(r_\parallel) \bigr]\\
      & + \Im(\alpha_{\rE,z}^\rE)\frac{\kappa^2}{k_0^2} \Im(r_\parallel) \\
      & + \Im(\alpha_{\rE,x}^\rH)
        \bigl[ \Im(r_\parallel) + \frac{\gamma^2}{k_0^2} \Im(r_\perp) \bigr]\\
      & + \Im(\alpha_{\rE,z}^\rH)\frac{\kappa^2}{k_0^2} \Im(r_\perp) 
	\biggr\} \; .
\end{split}
\label{Eq:Sample_to_Ellipsoid_ev}
\end{equation} 
Again this result leads directly to the corresponding expression for a 
sphere~\cite{ChapuisEtAl08}:
\begin{equation}
\begin{split}
	\langle P_{\rev, \rS \rightarrow \rP} \rangle & = 
	\int_0^\infty\!\!\! \rd\omega\, \Theta(\omega, T_\rS) 
         \, k_0^2 \!\!\!\int\limits_{\kappa > k_0}\!\!\! \frac{\rd^2 \kappa}{(2 \pi)^2} 
	\frac{\re^{-2 \gamma a}}{2 \gamma} \biggl\{ \\
      & \Im(\alpha_{\rP}^{\rE})
        \bigl[ \Im(r_\perp) + \frac{2 \kappa^2 - k_0^2}{k_0^2} \Im(r_\parallel)
        \bigr] \\
    + & \Im(\alpha_{\rP}^{\rH})
        \bigl[ \Im(r_\parallel) + \frac{2 \kappa^2 - k_0^2}{k_0^2} \Im(r_\perp)
	\bigr] \biggr\} \; .
\end{split}
\label{Eq:Sample_to_Sphere_ev}
\end{equation}
Note that in general one has  
$\varepsilon_0\langle |\mathbf{E}_\rf|^2 \rangle_\omega \neq 
\mu_0\langle |\mathbf{H}_\rf|^2 \rangle_\omega$ 
for the evanescent modes. In contrast to the transfer by propagating modes, 
this implies that the magnetic contribution to the heat transfer, which is 
proportional to $\Im(\alpha_{\rP}^{\rH})$, can dominate the electric 
one~\cite{ChapuisEtAl08} even if 
$\Im(\alpha_{\rP}^{\rE}) > \Im(\alpha_{\rP}^{\rH})$.

In the quasi-static regime, i.e., for distances $a$ even smaller than the
substrate skin depth, one has $\kappa \gg \sqrt{|\epsilon|} k_0$, so that 
Eq.~(\ref{Eq:Sample_to_Ellipsoid_ev}) leads to the approximation 
\begin{equation}
\begin{split}
  	\langle P_{\rev,\rS\rightarrow\rE} \rangle & \approx 
	\frac{2}{\pi}\int_0^\infty \!\!\! \rd\omega \, \Theta(\omega, T_\rS) 
        \, k_0^2\!\!\int_{0}^\infty \!\frac{\rd \kappa}{2 \pi}  
        \frac{\kappa^2 \re^{-2 \kappa a}}{k_0^2} \\
      & \times\biggl\{ 
        \Im(\alpha_{\rE,x}^\rE + \alpha_{\rE,z}^\rE) \Im(r_\parallel)\\
      & \quad+ \Im(\alpha_{\rE,x}^\rH + \alpha_{\rE,z}^\rH) \Im(r_\perp) \biggr\}
	\; .
\end{split}
\end{equation}
Furthermore, using the approximate expressions 
\begin{align}
  	\Im(r_\parallel) &\approx  
	\frac{2 \varepsilon_\rS''}{|\varepsilon_\rS + 1|^2} \; , \\
  	\Im(r_\perp)     &\approx 
	\frac{\varepsilon_\rS'' k_0^2}{ 4 \kappa^2}
\end{align}
for the imaginary parts of the reflection coefficients, the integration 
over the wave number $\kappa$ can be carried out, giving
\begin{equation}
\begin{split}
  	\langle P_{\rev,\rS\rightarrow\rE} \rangle & \approx 
	\frac{2}{\pi^2}\int_0^\infty \!\!\! \rd\omega\, \Theta(\omega, T_\rS) 
         \\
      & \times\biggl\{ 
        \Im(\alpha_{\rE,x}^\rE + \alpha_{\rE,z}^\rE) 
	\frac{\varepsilon_\rS''}{|\varepsilon_\rS + 1|^2} 
	\frac{1}{(2 a)^3}\\
      & \quad + \Im(\alpha_{\rE,x}^\rH + \alpha_{\rE,z}^\rH)
	\frac{k_0^2\varepsilon_\rS''}{32 a} \biggr\} \; .
\end{split}
\label{Eq:quasistatisch}
\end{equation}
As expected, in the quasi-static regime one has 
$\langle P^\rE_{\rev,\rS\rightarrow\rE}\rangle \propto a^{-3}$ 
for the electric and 
$\langle P^\rH_{\rev,\rS\rightarrow\rE} \rangle \propto a^{-1}$ for the 
magnetic contribution; these power laws are the same as those for a sphere.
Since the sums $\Im(\alpha_{\rE,x}^\rE + \alpha_{\rE,z}^\rE)$ and 
$\Im(\alpha_{\rE,x}^\rH + \alpha_{\rE,z}^\rH)$ appear in the frequency 
integral, one may expect that in this regime 
$\langle P_{\rev,\rS\rightarrow\rE}\rangle$ behaves like the imaginary part 
of the sum of the polarizabilities at the thermal frequency when $R_\rs$ and 
$R_\rp$ are varied.

It needs to be stressed, however, that this approximate 
result~(\ref{Eq:quasistatisch}) has to be taken with some caution, since 
the required short particle-sample distances may conflict with those
required by the dipole approximation.

\section{Discussion}
\label{Sec_Discus}

Now we study the shape dependence of the radiative heat transfer between a 
spheroidal nanoparticle and a planar sample numerically, on the basis of
the full expressions~(\ref{Eq:Sample_to_Ellipsoid_pr}) and  
(\ref{Eq:Sample_to_Ellipsoid_ev}). To this end, we assume the temperatures 
$T_\rP = 100\,{\rm K}$ for the particle and $T_\rS = 300\,{\rm K}$ for the
sample. The permittivity of the sample is described by the bulk Drude 
model~(\ref{Eq:Drude-Modell}) with parameters for gold; that of the particle
by the Drude expresson with the plasma frequency of gold, and with the
relaxation rate adapted to surface scattering, as before. Since we are 
interested in the variation of the heat transfer caused by alterations of the 
shape of the nanoparticle, we vary the ratio $R_\rs/R_\rp$ in such a way that 
the volume $V_\rE$ of the respective spheroidal nanoparticle equals the volume 
$V_\rP$ of a sphere with radius $R = 20$~nm. We estimate that the range 
$1/5 \le R_\rs/R_\rp \le 5$ is about the largest that is still compatible 
with the dipole approximation, and restrict ourselves to this interval.

\begin{figure}[Hhbt]
\epsfig{file=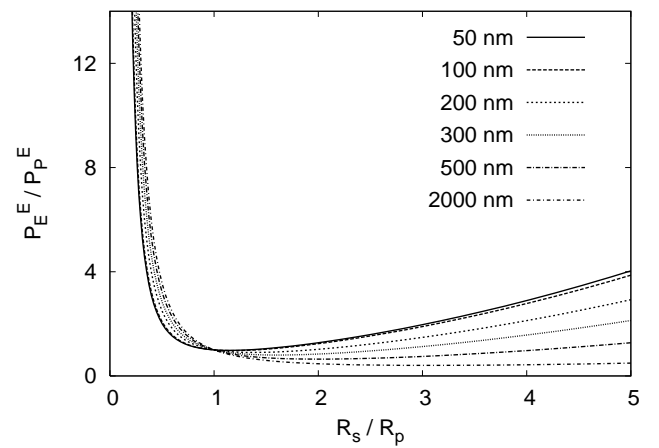, width=0.48\textwidth}
\caption{Electric contribution $P_\rE^\rE$ to the radiative heat transfer
	between a spheroidal gold particle with temperature  
	$T_\rP = 100\,{\rm K}$ and a planar gold surface with temperature
	$T_\rS = 300\,{\rm K}$, normalized to the corresponding value
	$P_\rP^\rE$ for a sphere. When the ratio $R_\rs/R_\rp$ is varied,
	the particle's volume is kept constant at that of a sphere with
	radius $R = 20$~nm. Line styles distinguish data for particle-sample
	distances ranging from $50\,{\rm nm}$ to $2000\,{\rm nm}$.}  
\label{Fig:OblatProlatZAbhaengigkeitPE}
\end{figure}

\begin{figure}[Hhbt]
\epsfig{file=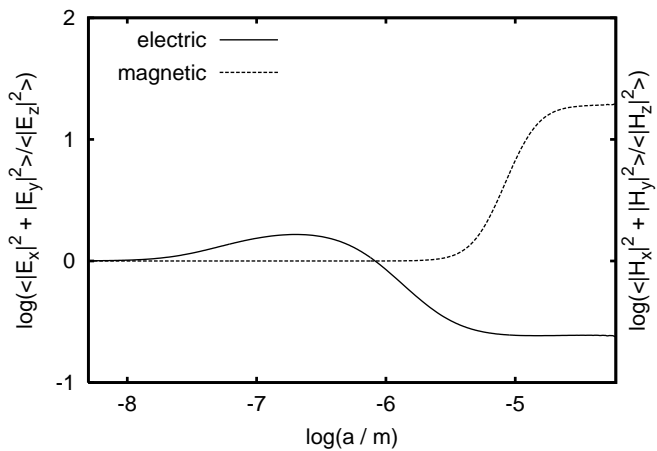, width=0.48\textwidth}
\caption{Ratios 
	$\langle |E_{x}|^2 + |E_{y}|^2 \rangle / \langle |E_{z}|^2 \rangle$ 
	(full line) and 
	$\langle |H_{x}|^2 + |H_{y}|^2 \rangle / \langle |H_{z}|^2 \rangle$ 
	(dashed line) above a semi-infinite planar gold sample 
	at $T_\rS = 300\,{\rm K}$, as function of the distance~$a$.}
\label{Fig:EparEperp}
\end{figure}

In Fig.~\ref{Fig:OblatProlatZAbhaengigkeitPE} we depict the electric
contribution to the total heat transfer $P_\rE^\rE$ vs.\ $R_\rs/R_\rp$ 
for several distances~$a$ between $50$~nm and $2000$~nm, normalized to the 
corresponding energy transfer $P_\rP^\rE$ between the isochoric sphere and 
the sample. Evidently, for small distances the curves of $P_\rE^\rE/P_\rP^\rE$  
actually resemble those of $\Im(\alpha_{\rE,x}^\rE + \alpha_{\rE,z}^\rE)$ 
in Fig.~\ref{Fig:Alpha_el}, as may have been conjectured from the 
quasi-static approximation~(\ref{Eq:quasistatisch}), notwithstanding its 
somewhat shaky justification. On the other hand, for relatively large 
distances the progression of $P_\rE^\rE/P_\rP^\rE$ is similar to that of 
$\Im(\alpha_{\rE,z}^\rE)$ alone. This behavior stems from the fact that in 
the latter regime, i.e., at distances such that the propagating modes 
dominate the heat transfer, one has $\langle |E_{z,\rpr}|^2 \rangle \gg 
\langle |E_{x,\rpr}|^2 + |E_{y,\rpr}|^2 \rangle$, as documented in 
Fig.~\ref{Fig:EparEperp}. In contrast, in the near-field regime where the 
evanescent modes dominate the transfer one finds 
$\langle |E_{z,\rev}|^2 \rangle = \langle |E_{x,\rev}|^2 
+ |E_{y,\rev}|^2 \rangle$, so that $\Im(\alpha_{\rE,x}^\rE)$ and 
$\Im(\alpha_{\rE,z}^\rE)$ contribute in equal measure there.   

\begin{figure}[Hhbt]
\epsfig{file=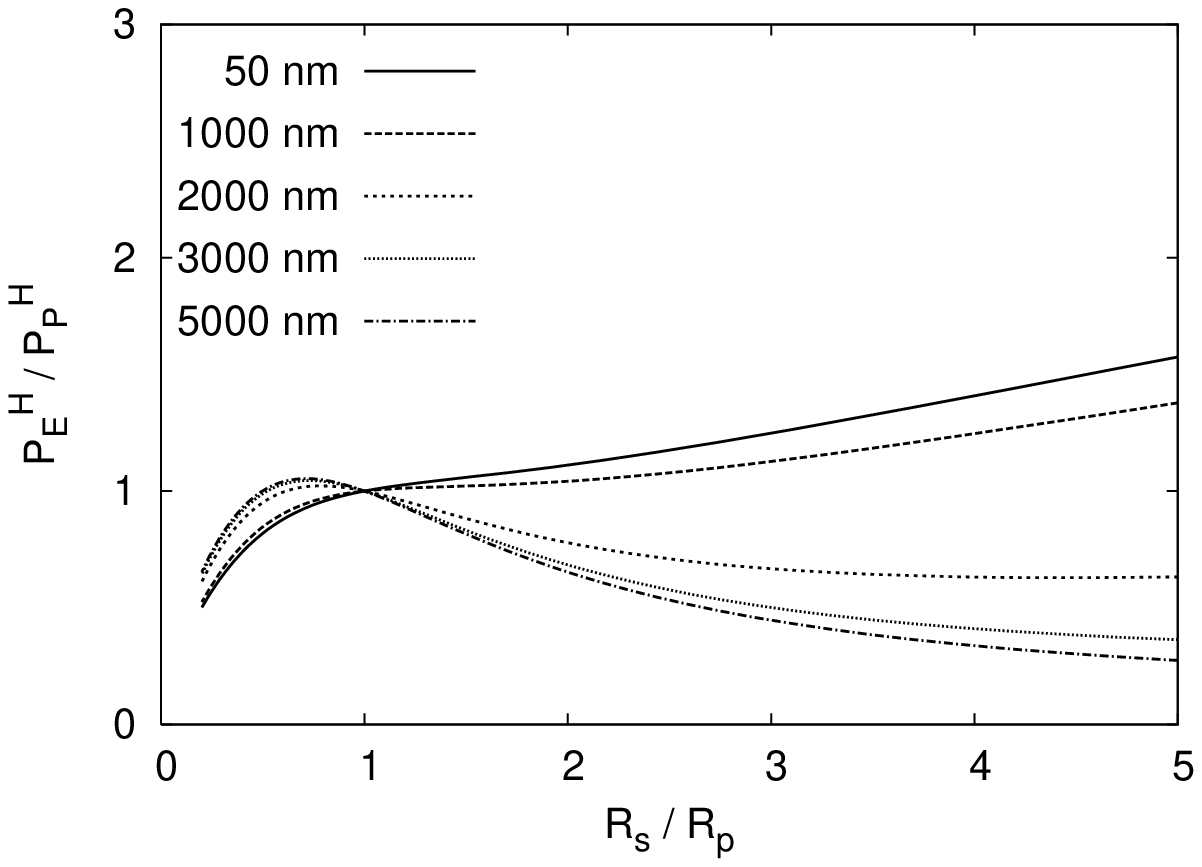, width=0.48\textwidth}
\caption{Magnetic contribution $P_\rE^\rH$ to the radiative heat transfer
	between a spheroidal gold particle with temperature  
	$T_\rP = 100\,{\rm K}$ and a planar gold surface with temperature
	$T_\rS = 300\,{\rm K}$, normalized to the corresponding value
	$P_\rP^\rH$ for a sphere. When the ratio $R_\rs/R_\rp$ is varied,
	the particle's volume is kept constant at that of a sphere with
	radius $R = 20$~nm. Line styles distinguish data for particle-sample
	distances ranging from $50\,{\rm nm}$ to $5000\,{\rm nm}$.}  
\label{Fig:OblatProlatZAbhaengigkeitPH}
\end{figure}

Analogously, we plot in Fig.~\ref{Fig:OblatProlatZAbhaengigkeitPH} the 
normalized magnetic contribution $P_\rE^\rH/P_\rP^\rH$ to the total heat 
transfer, for distances~$a$ extending from $50$~nm up to $5000$~nm. Similar 
to the electric case, in the near-field regime the graphs of 
$P_\rE^\rH/P_\rP^\rH$ resemble that of 
$\Im(\alpha_{\rE,x}^\rH + \alpha_{\rE,z}^\rH)$ previously shown in 
Fig.~\ref{Fig:Alpha_mg}, since $\langle |H_{z,\rev}|^2 \rangle = 
\langle |H_{x,\rev}|^2 + |H_{y,\rev}|^2 \rangle$ for small distances 
(see Fig.~\ref{Fig:EparEperp}). On the other hand, for relatively large
distances~$a$ the plots of $P_\rE^\rH/P_\rP^\rH$ are quite similar to
that of $\Im(\alpha_{\rE,x}^\rH)$ alone, since then 
$\langle |H_{z,\rpr}|^2 \rangle \ll  
\langle |H_{x,\rpr}|^2 + |H_{y,\rpr}|^2 \rangle$. 

\begin{figure}[Hhbt]
\epsfig{file=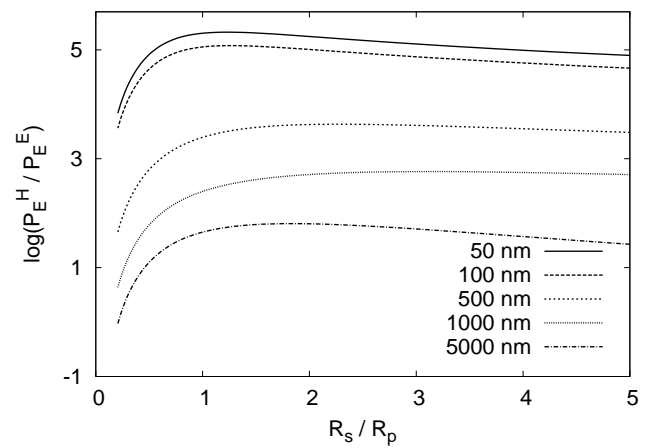, width=0.48\textwidth}
\caption{Logarithm of the ratio $P_\rE^\rH/P_\rE^\rE$ of the magnetic to 
	the electric contribution to the heat transfer vs.\ $R_\rs/R_\rp$,
	for the situation considered in 
	Figs.~\ref{Fig:OblatProlatZAbhaengigkeitPE}
	and~\ref{Fig:OblatProlatZAbhaengigkeitPH}.}
\label{Fig:OblatProlatZAbhaengigkeitPEPH}
\end{figure}

The ratio of the magnetic to the electric contribution $P_\rE^\rH/P_\rE^\rE$ 
is drawn in Fig.~\ref{Fig:OblatProlatZAbhaengigkeitPEPH}, again for distances 
from $50$~nm to $5000$~nm. For all distances considered the magnetic 
contribution to the heat transfer is much greater than the electric one, 
unless $R_\rs$ is much smaller than $R_\rp$. This fact has been discussed in 
detail by Chapuis {\itshape et al.}~\cite{ChapuisEtAl08} for a spherical 
metallic nanoparticle above a planar metallic sample. Here we find that the 
curves of the ratio $P_\rE^\rH/P_\rE^\rE$ exhibit a maximum near 
$R_\rs/R_\rp = 1$, so that the dominance of the magnetic part is somewhat less 
pronounced for nanoparticles with shapes differing from a sphere. For markedly 
needle-like, prolate particles with a very small ratio $R_\rs/R_\rp$, the 
graphs suggest that the electric contribution could even dominate the
magnetic one, because the induction of Foucault currents would be suppressed 
by the needle-like shape. We remark that for non-metallic nanoparticles and 
non-metallic samples the contribution due to induced eddy currents can be 
neglected anyway, so that for such materials it is always the electric 
contribution which dominates the heat transfer.    

We conclude from Figs.~\ref{Fig:OblatProlatZAbhaengigkeitPE} and 
\ref{Fig:OblatProlatZAbhaengigkeitPH} that the near-field heat transfer 
between a metallic nanoparticle and a sample depends to a sizeable extent 
on the nanoparticle's shape even when the total radiating volume is kept 
constant. 
In the example considered, the electric contribution to the heat transfer 
is enhanced for strongly prolate spheroids by a factor of about 10 as
compared to a perfect sphere, and by a factor of about 4 for strongly oblate 
ones. On the other hand, the magnetic contribution is substantially reduced 
in the needle-like case, but exceeds the sphere value by a factor of about 2 
for pancake-like particles, always assuming the orientation specified in 
Fig.~\ref{Fig:Ellipsoid}.

\section{Conclusions}

We have investigated the radiative heat transfer between a metallic spheroidal
nanoparticle and a planar metallic probe for small and large distances, on 
the basis of the analytical expressions~(\ref{Eq:Sample_to_Ellipsoid_pr}) 
and (\ref{Eq:Sample_to_Ellipsoid_ev}) for the heat transfer mediated by 
propagating and evanescent modes, respectively. By numerical analysis for the 
example of a spheroidal gold particle above a gold surface we have investigated
in detail the precise shape dependence of the radiative heat transfer in both 
the near-field and the far-field regime. Furthermore, we have derived the 
approximate expression~(\ref{Eq:quasistatisch}) for the nonretarded evanescent 
regime, which provides a qualitative understanding of the numerical results. 

\begin{figure}[Hhbt]
\epsfig{file=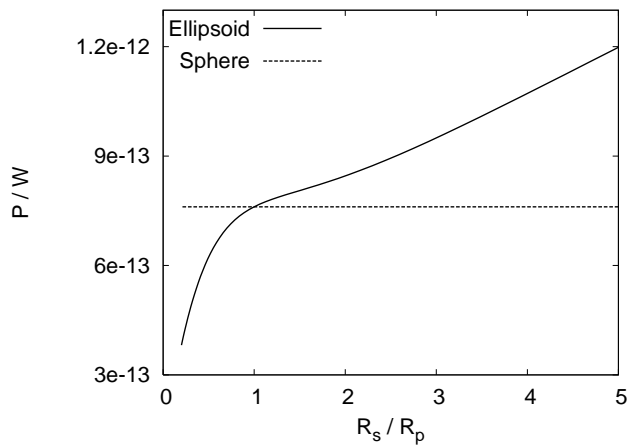, width=0.48\textwidth}
\caption{Heat transfer between a gold spheroid with temperature 
	$T_\rP = 100\,{\rm K}$ and a planar gold surface with temperature
	$T_\rS = 300\,{\rm K}$ separated by $100$~nm. The volume of the
	spheroid is kept constant at that of a sphere with radius $R = 20$~nm
	when $R_\rs/R_\rp$ is varied. The horizontal line marks the value 
	of the heat current between the isochoric sphere and the sample.}
\label{Fig:absolut}
\end{figure}

Figure~\ref{Fig:absolut} shows the absolute total heat transfer between 
a gold spheroid with temperature $T_\rP = 100\,{\rm K}$ and a planar gold
surface with temperature $T_\rS = 300\,{\rm K}$ for the separation
$a = 100$~nm, again fixing the spheroid's volume at that of a sphere with 
radius $R = 20$~nm when changing the aspect ratio. For $R_\rs/R_\rp = 1/5$ 
the heat transfer between spheroid and sample is only about half of that 
between sphere and sample, whereas for $R_\rs/R_\rp = 5$ the spheroid-sample 
transfer is roughly two times as efficient as that occurring in the 
sphere-sample geometry. The very fact that there exists such a marked shape 
dependence might be of interest for nanoscale thermal engineering, insofar as 
it appears possible to control the amount of heat transported at nanoscale 
distances by carefully designing the shapes of both the emitting and the 
absorbing pieces. 

Our discussion is subject to several restrictions: There is the dipole 
approximation~(\ref{Eq:P_Sample_to_Probe_general}) made right at the outset, 
the assumption of constant fields $\mathbf{H}_\rf$ inside the particle 
entering into Eq.~(\ref{Eq:E_H_Ellipsoid}), and the simplified 
correction~(\ref{Eq:SurfScat}) for surface scattering. Taken together, 
these simplifications render our analysis approximate, rather than exact, 
although they should still capture the essential physics. Further issues 
that should be investigated in future works concern the possible effects 
of spatial dispersion~\cite{ChapuisEtAl08c}, and of surface 
roughness~\cite{BiehsGreffet09}.   

With respect to the problem of quantifying the actual heat transfer in
a near-field scanning thermal microscope, our study helps to pin down the
error margin. A typical NSThM sensor tip is larger than the nanoparticles 
considered in our examples, and does possess an internal structure, but
again the precise shape of the sensor will influence the heat current 
it records. In view of our model calculations, we estimate that the values 
obtained on the grounds of the dipole model under the assumption of an ideal 
spherical sensor may deviate, in the appropriate distance regime, by up to 
an order of magnitude from the true values, but probably not by more. 

\section{Acknowledgments}

This work was supported by the Deutsche Forschungsgemeinschaft through 
Grant No.\ KI 438/8-1. S.-A.~B.\ gratefully acknowledges a fellowship 
awarded by the Deutsche Akademie der Naturforscher Leopoldina under 
Grant No.\ LPDS 2009-7.

\end{document}